\documentclass[prd,nofootinbib,showpacs,preprintnumbers,amssymb]{revtex4} 
\usepackage{graphicx}
\usepackage{epsfig}
\usepackage{dcolumn}
\usepackage{bm}
\usepackage{amsmath}
\def\gsim{\lower0.5ex\hbox{$\:\buildrel >\over\sim\:$}}
\def\lsim{\lower0.5ex\hbox{$\:\buildrel <\over\sim\:$}}

\newcommand{\be}{\begin{equation}}
\newcommand{\ee}{\end{equation}}
\newcommand{\bea}{\begin{eqnarray}}
\newcommand{\eea}{\end{eqnarray}}

\newcommand{\nbox}{{\,\lower0.9pt\vbox{\hrule \hbox{\vrule height 0.2 cm
\hskip 0.2 cm \vrule height 0.2 cm}\hrule}\,}}

\begin{document}

\preprint{UCI-TR-2009-05}

\title{Parameter Space of General Gauge Mediation}

\author{Arvind Rajaraman$^a$}
\email{arajaram@uci.edu}
\author{Yuri Shirman$^a$}
\email{yshirman@uci.edu}
\author{Joseph Smidt$^a$}
\email{jsmidt@uci.edu}
\author{Felix Yu$^a$}
\email{felixy@uci.edu}
\affiliation{$^a$Department of Physics and Astronomy, University of California,
Irvine, CA 92697, USA}

\date{\today}

\begin{abstract}

We study a subspace of General Gauge Mediation (GGM) models which
generalize models of gauge mediation.  We find superpartner spectra
that are markedly different from those of typical gauge and gaugino
mediation scenarios.  While typical gauge mediation predictions of
either a neutralino or stau next-to-lightest supersymmetric particle
(NLSP) are easily reproducible with the GGM parameters, chargino and
sneutrino NLSPs are generic for many reasonable choices of GGM
parameters.
\end{abstract}


\maketitle

\section{Introduction}
Supersymmetry (SUSY) is one of the most well-motivated and
well-studied extensions to the Standard Model (SM).  If SUSY is broken
at about the TeV scale, it solves the gauge hierarchy problem of the
SM and also explains the existence of dark matter.  One major problem
with such models is the presence of flavor changing neutral currents
(FCNCs); this has motivated the study of gauge mediated supersymmetry
breaking (GMSB) where SUSY breaking is communicated to the observable
sector through the Standard Model interactions
\cite{bib:DFS,bib:DF,bib:DR81,bib:DN,bib:DR83,bib:A-GCW,bib:NO,bib:DNNS,bib:DNS,bib:GR}.
Flavor violating dynamics are then available only through Yukawa
interactions; any new FCNCs are aligned with the SM and are therefore
small.

From the phenomenological perspective, GMSB is usually described
through the introduction of heavy messenger fields. These models are
characterized by two parameters: the messenger scale $M$, where the
soft parameters are generated, and $\Lambda$ the effective scale of
SUSY breaking in the visible sector. Up to renormalization effects,
the superpartner masses are then completely determined\footnote{We do
not include $\mu$ and $B$ parameters in our analysis.} by their
Standard Model quantum numbers and $\Lambda$.  Typical weakly coupled
GMSB models are quite complicated, however, and require a relatively
high SUSY breaking scale.  It would be exciting if SUSY were broken at
low energies so that messenger fields or even the hidden SUSY breaking
sector would be directly observable in experiment.

For this reason, significant effort has been devoted over the years to
the search for models of low energy direct gauge mediation (see for
example \cite{Izawa:1997gs,Csaki:2006wi,Dine:2007dz}) or even single
sector SUSY breaking \cite{Terning:1999at}.  It is unfortunately quite
non-trivial to calculate the superpartner spectrum in the strongly
interacting theories needed for implementation of direct and/or low
energy gauge mediation. Indeed, it was shown in \cite{bib:CRS,bib:RS}
that usually neglected renormalization effects from a strongly
interacting hidden sector may radically modify standard GMSB
predictions.

Recently, Meade, Seiberg, and Shih \cite{bib:MSS} have succeeded in
giving a general characterization of spectra in gauge mediated models,
including strongly interacting ones.  While certain GMSB features such
as gravitino LSP, smallness of $A$-terms and certain sum rules remain
unchanged, their results imply that spectra which are quite different
from those of traditional GMSB can be obtained. Examples of weakly
interacting models of this type were presented in \cite{bib:CDFM} and
further generalized in \cite{Buican:2008ws}.\footnote{See
\cite{bib:IS,bib:Carpenter,bib:SVW,bib:CFNV} for other related work.}

Our goal in this paper is to begin the study of the phenomenology of
the GGM scenario. These studies are likely to constrain the allowed
GGM parameter space and also suggest new experimental signatures which
do not arise in minimal GMSB models. In particular, we shall attempt
to construct viable models with low messenger scales, which may allow
us to probe the hidden sector directly.

We will begin by reviewing the general gauge mediation framework in
section \ref{sec:GGM}.  In section \ref{sec:mGGMfmwk}, we begin the
study of the phenomenology of GGM by considering models where sfermion
and gaugino masses are controlled by two independent parameters, and
the overall scale of the theory is the third parameter (a different
slice of GGM parameter space was recently studied in
\cite{bib:Carpenter}). We then demonstrate that even our simplified
subset of GGM parameters can be used to create models with new and
possibly interesting phenomenology, and we qualitatively detail how
these new mass hierarchies differ from previously considered models of
GMSB and gaugino mediation models.  We conclude by calculating the
mass spectrum at certain benchmark points and showing their
qualitative phenomenological differences from previously considered
gauge mediated models.

\section{Review of General Gauge Mediation}
\label{sec:GGM}

To describe strongly interacting models of GMSB, a model independent
formulation of gauge mediation is necessary. Such a formulation was
proposed in \cite{bib:MSS}: the theory decouples into the MSSM and a
separate SUSY breaking sector in the limit where MSSM gauge couplings
tend to zero (and $M_{\mathrm{Pl}}$ to infinity). With these
assumptions, one may relate the superpartner spectrum to one- and
two-point correlation functions of the supercurrent
$\mathcal{J}$. Current conservation $D^2 \mathcal{J} = \overline{D}^2
\mathcal{J} = 0$ leads to the following expressions for one- and
two-point correlation functions\footnote{See \cite{bib:MSS} and
\cite{bib:IS} for more details.}:
\begin{eqnarray}
&&\langle J (x) \rangle = \zeta \nonumber\\
&&\langle J (p) J (-p) \rangle = \tilde{C}_0 (p^2 / M^2;
  M / \Lambda_{UV})\nonumber \\
&&\langle j_{\alpha} (p) \overline{j}_{\dot{\alpha}} (-p) \rangle =
-\sigma^{\mu}_{\alpha \dot{\alpha}} p_{\mu} \tilde{C}_{1/2}
(p^2 / M^2; M / \Lambda_{UV}) \\
&&\langle j_{\mu} (p) j_{\nu} (p) \rangle = - (p^2 \eta_{\mu \nu} -
p_{\mu} p_{\nu} ) \tilde{C}_1 (p^2 / M^2; M / \Lambda_{UV} )
\nonumber\\
&&\langle j_{\alpha} (p) j_{\beta} (-p) \rangle = \epsilon_{\alpha
\beta} M \tilde{B}_{1/2} (p^2 / M^2)\nonumber
\end{eqnarray}
where $M$ is a characteristic scale of the theory, $\Lambda_{UV}$ is a
UV cutoff and a common factor of $(2 \pi)^4 \delta^{(4)} (0)$ is
understood. The four functions $\tilde{C}_0$, $\tilde{C}_{1/2}$,
$\tilde{C}_1$, and $\tilde{B}_{1/2}$ serve to characterize the hidden
sector contribution to the current-current correlators.

When these currents carry MSSM quantum numbers, each $\tilde{C}_{j =
0, 1/2, 1}^{(r)}$ and $\tilde{B}_{1/2}^{(r)}$ gains a new index $r =
3, 2, 1$ which labels the SU(3) $\times$ SU(2) $\times$ U(1) gauge
groups.  After considering the effective action from this coupling of
hidden sector correlators with the gauge supermultiplets, the gaugino
and sfermion masses are given in the effective theory at scale $M$ by
\begin{equation}\begin{array}{l}
M_r = g_r^2 M \tilde{B}_{1/2}^{(r)} (0) \\
m_{\tilde{f}}^2 = g_1^2 Y_f \zeta + \sum\limits_{r = 1}^3 g_r^4 c_2
(f; r) M^2 A_r,
\label{eqn:masses}
\end{array}\end{equation}
where
\begin{equation}
\begin{array}{lll}
A_r &=& - \int \frac{d^4 p}{(2 \pi)^4} \frac{1}{M^2 p^2} \left(
3 \tilde{C}_1^{(r)} (p^2 / M^2) - 4 \tilde{C}_{1/2}^{(r)} (p^2 / M^2)
+ \tilde{C}_0^{(r)} (p^2 / M^2) \right) \\
 &=& - \frac{1}{16 \pi^2} \int dy \left( 3 \tilde{C}_1^{(r)} (y)
 - 4 \tilde{C}_{1/2}^{(r)} (y) + \tilde{C}_0^{(r)} (y) \right) \\
\end{array}
\label{eqn:Adef}
\end{equation}
and $c_2 (f;r)$ is the quadratic Casimir invariant of gauge group $r$
of fermion $f$.  Since superpartner masses in (\ref{eqn:masses}) are
generated at the messenger scale, it is convenient to identify the
scale $M$ with the scale of messenger masses.  We also note that in
addition to its dependence on $A_r$, $B_r$ and $\zeta$, the
superpartner spectrum at the electroweak symmetry breaking scale is
modified due to renormalization group (RG) evolution between the
messenger scale $M$ and the electroweak scale.

We would now like to identify benchmark points in the parameter space
of GGM.  To this end we will consider GMSB models with $N_5$
messengers in the $5$ and $\overline{5}$ representations of $SU(5)$.
As usual, fermionic components of messenger supermultiplets are
characterized by mass $M$ while scalar components have mass squared
$M^2\pm F$.  In regular GMSB models \cite{Martin:1996zb},
\begin{equation}
\begin{array}{l}
M_r = \frac{\alpha_r}{4 \pi} N_5 \frac{F}{M} g(x)
\\
m_{\tilde{f}}^2 = 2 N_5 \left| \frac{F}{M} \right|^2 \sum\limits_r 
\left( \frac{\alpha_r}{4 \pi} \right)^2 c_2 (f; r) f(x),
\label{eqn:GMSBgenmasses}
\end{array}
\end{equation}
where
\begin{equation}\begin{array}{l}
g(x) = \left. \frac{1}{x^2} \right.
\left[ (1+x) \log (1+x) + (1-x) \log (1-x) \right] 	\\
f(x) = \left. \frac{1+x}{x^2} \right.
\left[ \log (1+x) - 2 \text{Li}_2 (x/[1+x]) +
\frac{1}{2} \text{Li}_2 (2x / [1+x]) \right] + (x \rightarrow -x),
\\
\label{eqn:GMSBgenmassesfg}
\end{array}
\end{equation}
where $x = F/M^2$.

\begin{figure}
\includegraphics[width=0.5\linewidth]{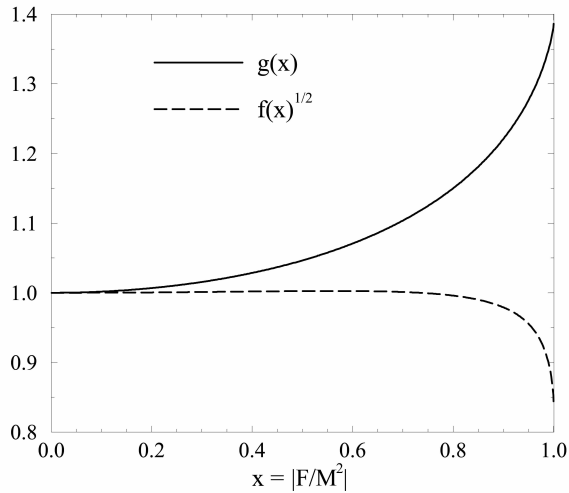}
\caption{The functions $g(x)$ and $\sqrt{f(x)}$ parameterizing gaugino
and sfermion masses, respectively. Figure taken from
\cite{Martin:1996zb}. }
\label{fig:fg}
\end{figure}

For small $x$, $g(x)$ and $f(x)$ both approach 1, and we can specify
points in the GGM parameter space corresponding to models of minimal
GMSB:
\begin{equation}\begin{array}{l}
B_r = B = \frac{N_5}{16 \pi^2} \left( \frac{F}{M^2} \right) \\
A_r = A = 2 \frac{N_5}{(16 \pi^2)^2} \left( \frac{F}{M^2} \right)^2 
= 2 \frac{B^2}{N_5} \\
\zeta = 0\,.\label{eqn:BAcorrelation}
\end{array}
\end{equation}

Since soft masses are generated at one loop, it is often convenient to
discuss the low energy spectrum in terms of the effective SUSY
breaking scale in the visible sector, $\Lambda = F / M \sim 100$
TeV. In GGM, on the other hand, the effective scale of SUSY breaking
could be smaller than $10^5$ GeV since gauginos formally arise at tree
level and sfermion mass squareds arise at one loop.

\section{The Minimal General Gauge Mediation Framework}
\label{sec:mGGMfmwk}

In this paper we will study a three parameter subspace of GGM
models. We will assume that $\zeta = 0$ and that values of $A_r$ and
$B_r$ are independent of the gauge group.  These models are described
by a characteristic messenger scale $M$, an overall suppression of
gaugino and sfermion masses relative to $M$, and the ratio between
these masses.

Note that in minimal GMSB models, gaugino masses arise at 1-loop and
sfermion mass squareds arise at 2-loops in the effective theory:
hence, gaugino and sfermion masses naturally both have a 1-loop
suppression factor and are approximately equal at the messenger scale.
One can see from (\ref{eqn:GMSBgenmasses}) and
(\ref{eqn:GMSBgenmassesfg}) that the ratio of gaugino and sfermion
soft masses can be modified in models with $F/M^2 \sim 1$; however,
this typically happens in models with a strongly interacting dynamical
SUSY breaking sector and a messenger scale close to 100 TeV.  While
calculable examples exist in the literature
\cite{Izawa:1997gs,Csaki:2006wi}, they typically allow only
$\mathcal{O}(1)$ changes in sfermion to gaugino mass ratios. In
effect, there is very little freedom in GMSB models to change the
ratio of sfermion mass to gaugino mass by more than an
$\mathcal{O}(1)$ factor.  Our three parameters allow us to explore the
consequences of tuning the sfermion to gaugino mass ratio by a large
factor.

An alternative approach to modifying superpartner mass ratios would
involve allowing the number of messengers, $N_5$, to take arbitrary
values. This approach was considered in Ref. \cite{Martin:1996zb} by
adopting messengers to be in extended gauge messenger multiplets.
Clearly, since gaugino and sfermion masses scale as $N_5$ and
$\sqrt{N_5}$, respectively, an arbitrary mass ratio can be obtained if
$N_5$ were allowed to be arbitrarily large.  Indeed, GMSB models with
extremely large values of $N_5$ are interesting since the superpartner
spectrum in this limit is the spectrum of gaugino mediation
($\tilde{g}$MSB) \cite{bib:CLNP,bib:KKS,bib:CKSS}.  Typically, only
small values of $N_5 < 5$ to $10$ (depending on the messenger scale) are
considered in the literature in order to maintain perturbativity.

Our parametrization allows the interpolation between spectra of
minimal GMSB and those of $\tilde{g}$MSB.  In fact, it is more general
than either of these two scenarios. This is due to the fact that in
$\tilde{g}$MSB models considered so far, the compactification
(messenger) scale was large; hence, as a consequence of RG evolution,
at the electroweak scale sfermion masses become comparable to gaugino
masses. In contrast, GGM allows the messenger scale to be as low as 10
TeV which results in a ``pure'' $\tilde{g}$MSB spectrum with sfermions
significantly lighter than gauginos.

\begin{figure}[h!]
\includegraphics[width=12cm]{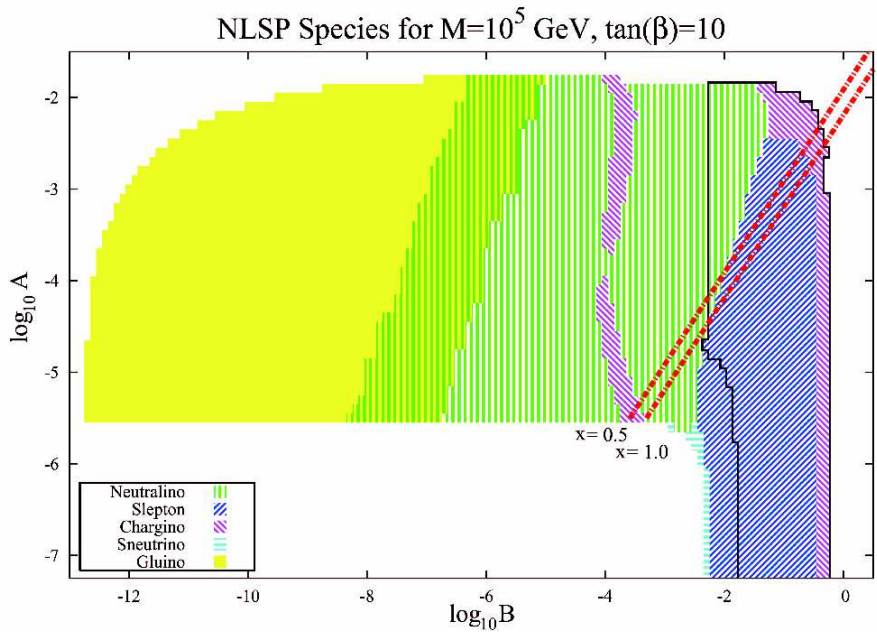}
\caption{(color online). NLSP regions for $M = 10^5$ GeV, $\tan \beta
= 10$, in general gauge mediation.  The solid outlined region
satisfies constraints on NLSP mass from direct experimental searches.
The dashed lines indicate the equivalent $B$ and $A$ relations for
$x = 0.5$ and $1.0$ in (\ref{eqn:BAcorrelation}).}
\label{fig:M5species}
\end{figure}

\begin{figure}[h!]
\includegraphics[width=12cm]{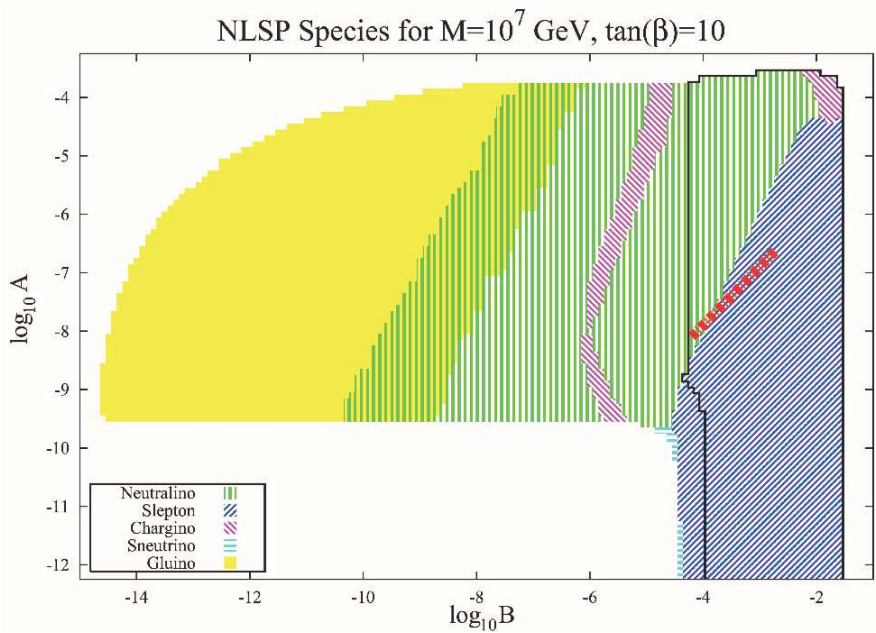}
\caption{(color online). Same as Fig.~(\ref{fig:M5species}), except
with $M = 10^7$ GeV.  The dashed line show models equivalent to GMSB
models with $\Lambda = 10^5$ GeV, $M_{\text{mess}} = 10^7$ GeV, and
$N_5 = 1$ to $30$.}
\label{fig:M7species}
\end{figure}

In our calculation, we will assume that the mass couplings $m_{H_1}^2$
and $m_{H_2}^2$ for both Higgs doublets are equal to the left-handed
slepton soft mass squared term $m_{\tilde{L}_L}^2$ at the scale $M$.
We set $\tan \beta = 10$ and we set the gravitino mass to be $m_{3/2}
= F / (\sqrt{3} M_{\text{Pl}})$, where $F/M = 10^5$ GeV and
$M_{\text{Pl}} = (8 \pi G_N)^{-1/2} \simeq 2.4 \times 10^{18}$ GeV is
the reduced Planck scale\footnote{We performed some scans varying
$\tan \beta$ from 3 to 50 and observed qualitatively similar
results.}.  We then use Softsusy 2.0.17 \cite{bib:Softsusy} to
determine the superpartner spectra.  Parameter regions (which we
require to have a ground state with broken electroweak symmetry) and
the corresponding NLSP species are shown in Fig.~(\ref{fig:M5species})
and Fig.~(\ref{fig:M7species}) for different choices of the messenger
scale.  We also impose the current experimental lower mass bounds on
NLSPs in GMSB scenarios: neutralinos at 114.0 GeV, squarks at 250.0
GeV, sleptons at 87.4 GeV, charginos at 101.0 GeV, sneutrinos at 43.7
GeV \cite{bib:PDG}, and gluinos 240.0 GeV \cite{bib:BCG}.  These
bounds remove the gluinos and sneutrinos, but leave some of the
chargino region intact, as evident by the delineated regions in
Fig.~(\ref{fig:M5species}) and Fig.~(\ref{fig:M7species}).  In addition,
the points toward the right, where $\log_{10} B \rightarrow 0$, are
disfavored from a theoretical perspective, since the NLSP masses at
$m_Z$ in this limit are $\mathcal{O}$(10 TeV) and lead to a ``little
hierarchy'' problem.

\subsection{Analysis}

We can compare our spectra to a benchmark GMSB scenario with $\Lambda
= 10^5$ GeV, $M_{\text{mess}} = 10^5$ or $10^7$ GeV, and $N_5$ small.
The dashed line of Fig.~(\ref{fig:M7species}) depicts models with
GMSB-like parameters of $\Lambda = 10^5$ GeV, $M_{\text{mess}} = 10^7$
GeV, $N_5$ from 1 to 30, and with $m_{\tilde{f}} / m_{\tilde{g}}
\simeq \mathcal{O}(1)$, since $x_i$ is small and $g(x_i) \approx
f(x_i) \approx 1$.  For models that preserve gauge coupling
perturbativity up to the GUT scale, $N_5$ is necessarily smaller.  In
the case of low energy SUSY breaking with $\Lambda\sim M_{mess}\sim
10^5~\mathrm{GeV}$ the ratio of corresponding GGM parameters $A$ and
$B$ is very sensitive to the value of $x=F/M^2$ and can vary by a
factor of 2. This is taken into account in Fig.~(\ref{fig:M5species}).

There are also regions where GGM parameters lead to spectra that are
markedly different than those of traditional GMSB models (Table 1). In
this table we show the spectra of three GGM models (which we will
refer to as GGM1, GGM2, and GGM3 respectively) and also two models of
traditional gauge mediation (which we will refer to as GMSB1, GMSB2
respectively).  In particular, we find that gauge mediation models do
not have to obey the typical low-energy hierarchy of
$m_{\tilde{\ell}_R} < m_{\tilde{\chi_1^0}} < m_{\tilde{\nu}_L} \simeq
m_{\tilde{\ell}_L} < m_{\tilde{\chi_2^0}} < m_{\tilde{q}}$ for the
case of a stau NLSP (cf. GMSB 1), or $m_{\tilde{\chi_1^0}} <
m_{\tilde{\ell}_R} < m_{\tilde{\nu}_L} \simeq m_{\tilde{\ell_L}} <
m_{\tilde{\chi_2^0}} < m_{\tilde{q}}$ in the case of a bino NLSP
(cf. GMSB 2).  For example, the model GGM1 has both gaugino masses and
sfermion mass squareds as approximately 1-loop suppressed relative to
the messenger scale $M = 10^5$ GeV, and leads to a SUSY spectrum with
all gauginos lighter than all scalars.  Model GGM2, on the other hand,
has gaugino masses approximately $1 / (128 \pi^3)$ suppressed, while
the sfermions start nearly massless, and all sleptons stay lighter
than all gauginos after RG evolution.  Clearly, the phenomenology of
such situations would be radically different from traditional GMSB
scenarios.  In the case of GGM1, squark production is heavily
suppressed at the LHC from kinematics and the parton momenta fraction,
and hence SUSY production is dominated by gluino pair production.  The
allowed modes for gluino decay are then a three-body final state of $q
+ q + \tilde{\chi}_i^0$ with 2 hard jets and missing energy (and a
likely $Z$ from $\tilde{\chi}_2^0 \rightarrow \tilde{\chi}_1^0$ decay)
or a two-body decay of $\tilde{g} \rightarrow g + G$.  On the other
hand, for GGM2, long cascade decay chains with lots of leptons are
possible.  These simple examples demonstrate two notably distinct sets
of signals, both originating from the general gauge mediation
framework.  We also present model GGM3, which sets both gaugino and
sfermion masses to be nearly tree-level in the theory, and because of
a large cancelation in the diagonalization of the gaugino mixing
matrices, the NLSP is a chargino.

\begin{table}
\caption{Comparison of Minimal General Gauge Mediation and
Minimal GMSB}
  \begin{tabular}{c|c||c|c|c||c|c|c}
 & & GGM1 & GGM2 & GGM3
  & & mGMSB1 & mGMSB2 \\
\hline
inputs: & $M$           & $10^5$ & $10^7$ & $10^4$
        & $M_{\text{mess}}$      & $10^5$   & $10^7$ \\
        & $\log_{10} B$ & -2.1   & -3.6   & -0.4
        & $\Lambda$              & $10^5$   & $10^5$ \\
        & $\log_{10} A$ & -2.25   & -10   & -1.5
        & $N_5$                  & 3        & 1 \\
        & $\tan\beta$ & 10     & 10     & 10
        &                        & 10       & 10 \\
        & $c_{\text{grav}}$    & 1.0 & 1.0 & 1.0
        &                        & 1.0      & 1.0 \\
        & sign$(\mu)$ & +      & +      & +
        &                        & +        & + \\
\hline
neutralinos: & $m_{\chi_1^0}$ & 174 & 533 & 137
             & & 547 & 132 \\
             & $m_{\chi_2^0}$ & 353 & 742 & 144
             & & 611 & 256 \\
             & $m_{\chi_3^0}$ & 1310 & 753 & 865
             & & 625 & 511 \\
             & $m_{\chi_4^0}$ & 1320 & 1030 & 1620
             & & 1080 & 522 \\
\hline
charginos: & $m_{\chi_1^{\pm}}$ & 345 & 736 & 128
           & & 603 & 258 \\
           & $m_{\chi_2^{\pm}}$ & 1310 & 1030 & 1630
           & & 1080 & 519 \\
\hline
Higgs: & $m_{h^0}$     & 123 & 116 & 116
       & & 116  & 112 \\
       & $m_{H^0}$     & 3050 & 849 & 735
       & & 833 & 619 \\
       & $m_{A^0}$     & 3050 & 849 & 734
       & & 833 & 619 \\
       & $m_{H^{\pm}}$ & 3050 & 853 & 739
       & & 837 & 624 \\
\hline
sleptons: & $m_{\tilde{\tau}_R}$ & 1390 & 174 & 348
          & & 276 & 191 \\
          & $m_{\tilde{e}_R}$    & 1400 & 179 & 349
          & & 279 & 195 \\
          & $m_{\tilde{\ell}_L}$ & 2770 & 417 & 728
          & & 578 & 372 \\
          & $m_{\tilde{\nu}_L}$  & 2780 & 410 & 727
          & & 574 & 363 \\
\hline
squarks: & $m_{\tilde{t}_1}$ & 7620 & 1620 & 2360
         & & 1860 & 875 \\
         & $m_{\tilde{t}_2}$ & 8260 & 1770 & 2500
         & & 2000 & 1010 \\
         & $m_{\tilde{u}_L}$ & 8390 & 1800 & 2510
         & & 2040 & 1050 \\
         & $m_{\tilde{u}_R}$ & 8070 & 1780 & 2440
         & & 1990 & 1010 \\
         & $m_{\tilde{d}_L}$ & 8430 & 1810 & 2520
         & & 1970 & 1060 \\
         & $m_{\tilde{d}_R}$ & 8030 & 1760 & 2430
         & & 2050 & 1000 \\
\hline
gluino: & $M_3$ & 1060 & 2630 & 4020
        & & 2740 & 804 \\
\hline
\end{tabular}
\end{table}

As pointed out in \cite{bib:MSS} another interesting feature of GGM
parameter space is that it interpolates between the phenomenology of
GMSB and $\tilde{g}$MSB models. In fact, GGM admits even more general
phenomenology. Indeed, while existing models of $\tilde{g}$MSB
\cite{bib:CLNP,bib:CKSS} have a large hierarchy between sfermion and
and gaugino masses at the compactification scale (usually taken to be
close to the GUT scale), this hierarchy is washed out at the TeV scale
due to the effects of long RG evolution. As expected, GGM easily
reproduces the spectra of such models. On the other hand, the bottom-up
approach of GGM allows one to take an effective messenger scale to be
as low as $10^4$ GeV leading to a ``pure $\tilde{g}$MSB'' spectrum at
the electroweak scale with the $4\pi$ suppression factor between
sfermion and gaugino masses intact.  The differences between low-scale
and high-scale $\tilde{g}$MSB are highlighted by a few representative
spectra in Table 2.  We note that because of the relatively small
running scale, low-scale $\tilde{g}$MSB models, like GGM4 and GGM5,
characteristically have slepton NLSPs, and hence a low-scale
$\tilde{g}$MSB model with a bino NLSP is disfavored.

\begin{table}

\caption{Gaugino mediation hierarchy comparison. The $\tilde{g}$MSB
hierarchies are taken from \cite{bib:CLNP}.  LEP constraints rule out
these models, but we present them for comparison.}
  \begin{tabular}{c|c||c|c|c||c|c|c|c}
 & & GGM4 & GGM5 & GGM6
 & & $\tilde{g}$MSB1 & $\tilde{g}$MSB2 & $\tilde{g}$MSB3 \\
\hline
inputs: & $M$              & $10^4$ & $10^5$ & $2 \times 10^{16}$
        & $m_{1/2}$   & 200       & 400       & 400 \\
        & $\log_{10} B$    & $-1.0$ & $-2.4$ & $-13.4$
        & $m_{H_u}^2$ & $(200)^2$ & $(400)^2$ & $(400)^2$ \\
        & $\log_{10} A$    & $-3.0$ & $-4.6$ & $-29$
        & $m_{H_d}^2$ & $(300)^2$ & $(600)^2$ & $(400)^2$ \\
        & $\tan \beta$& $10$   & $10$   & $10$
        & $\mu$       & 370       & 755       & 725 \\
        &             &        &        &
        & $B$         & 315       & 635       & 510 \\
        &             &        &        &
        & $y_t$       & 0.8       & 0.8       & 0.8 \\
\hline
neutralinos: & $m_{\chi_1^0}$ & 141 & 79  & 164
             & & 78  & 165 & 165 \\
             & $m_{\chi_2^0}$ & 172 & 144 & 308
             & & 140 & 315 & 315 \\
             & $m_{\chi_3^0}$ & 229 & 263 & 523
             & & 320 & 650 & 630 \\
             & $m_{\chi_4^0}$ & 430 & 291 & 537
             & & 360 & 670 & 650 \\
\hline
charginos: & $m_{\chi_1^{\pm}}$ & 158 & 143 & 311
           & & 140 & 315 & 315 \\
           & $m_{\chi_2^{\pm}}$ & 429 & 289 & 533
           & & 350 & 670 & 645 \\
\hline
Higgs: &               &     &     &
       & $\tan \beta$  & 2.5 & 2.5 & 2.5 \\
       & $m_{h^0}$     & 106 & 107 & 113
       & & 90  & 100  & 100 \\
       & $m_{H^0}$     & 220 & 315 & 579
       & & 490 & 995  & 860 \\
       & $m_{A^0}$     & 219 & 314 & 578
       & & 490 & 1000 & 860 \\
       & $m_{H^{\pm}}$ & 183 & 324 & 584
       & & 495 & 1000 & 860 \\
\hline
sleptons: & $m_{\tilde{e}_R}$    & 82  & 103 & 155
          & & 105 & 200 & 160 \\
          & $m_{\tilde{e}_L}$    & 163 & 206 & 281
          & & 140 & 275 & 285 \\
          & $m_{\tilde{\nu}_L}$  & 143 & 189 & 269
          & & 125 & 265 & 280 \\
\hline
stops: & $m_{\tilde{t}_1}$ & 616 & 600 & 653
       & & 350 & 685 & 690 \\
       & $m_{\tilde{t}_2}$ & 681 & 673 & 846
       & & 470 & 875 & 875 \\
\hline
other squarks: & $m_{\tilde{u}_L}$ & 657 & 666 & 856
               & & 470 & 945 & 945 \\
               & $m_{\tilde{u}_R}$ & 649 & 649 & 832
               & & 450 & 905 & 910 \\
               & $m_{\tilde{d}_L}$ & 665 & 674 & 862
               & & 475 & 950 & 945 \\
               & $m_{\tilde{d}_R}$ & 646 & 647 & 824
               & & 455 & 910 & 905 \\
\hline
gluino: & $M_3$ & 1135 & 536 & 938
        & & 520 & 1000 & 1050 \\
\hline
\end{tabular}
\end{table}

\begin{figure}[h!]
\includegraphics[width=12cm]{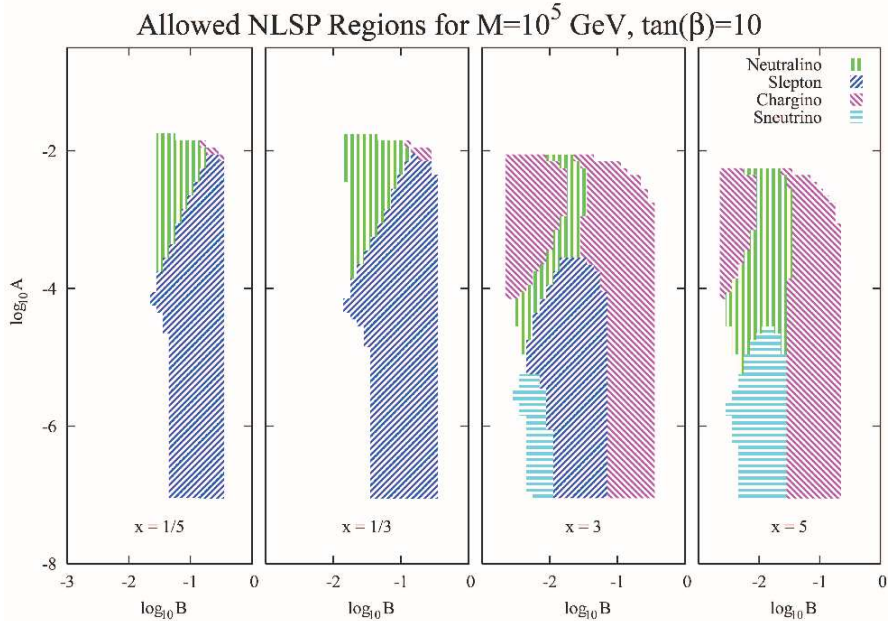}
\caption{(color online).  Allowed NLSP regions for $M = 10^5$ GeV,
$\tan \beta = 10$, in general gauge mediation.  Here, we set $A_1 = x
A_2 = x A_3 = x A$, and similarly for $B$, to use in
(\ref{eqn:masses}).}
\label{fig:HyperMulti}
\end{figure}

A notable feature evident in our GGM parameter plots is the
intersection of three discrete NLSP regions: neutralino (bino),
chargino (wino), and slepton (stau) NLSPs.  This triple point region
likely leads to very interesting phenomenology and new collider
signals, since all three particles are highly degenerate in this
parameter space.  In principle, there could also be other combinations
of NLSPs that give triple point phenomenology.  (For example, there is
a triple point of sneutrinos, neutralinos, and charginos in the
rightmost plot of Fig.~(\ref{fig:HyperMulti}).)  One immediate
consequence of highly degenerate NLSPs, however, is that the typical
dark matter relic density calculation must now include
coannihilations, and so a full analysis of triple point phenomenology
must first determine what the favored degenerate mass range is in
order to give an appropriate dark matter relic density.  We leave a
full analysis of triple point phenomenology for future work.

\subsection{Extensions of minimal set of parameters}

We now study how simple extensions of minimal set of parameters
considered so far affect our results by allowing the variation of
$A_1$ and $B_1$ relative to the other $A$ and $B$ parameters.
Specifically, we set $A_1 = x A_2 = x A_3 \equiv x A$ and $B_1 = x B_2
= x B_3 \equiv x B$, where $x = 1/5$, $1/3$, $3$, and $5$.

As can be seen in Fig.~(\ref{fig:HyperMulti}), this leads to
interesting phenomenological consequences.  In particular, note that
for $x > 1$ the charged slepton NLSP is not a generic prediction of
gauge mediation. Instead, for a large set of parameters, sneutrinos
and charginos become the NLSP. This is especially interesting in view
of the fact that the $x>1$ region of GGM parameter space squeezes the
superpartner spectrum and thus alleviates the little hierarchy
problem.

On a separate note, in Fig.~(\ref{fig:M5species}) and
Fig.~(\ref{fig:M7species}), there are new regions where gluino NLSPs
are present.  (In Fig.~(\ref{fig:M5species}), with $\log_{10} B$
between $-10$ and $-6$, the gluino region overlaps with the
neutralinos; similarly for Fig.~(\ref{fig:M7species}).)  The entire
gluino region has masses calculated at best to be $\mathcal{O}$ (0.1
GeV), and therefore ruled out; yet it should be possible to construct
GGM models with viable gluino NLSPs by adjusting the parameters $A_r$,
$\tilde{B}_{1/2}^{(r)}$ appropriately.

\section{Acknowledgements}

This work is supported in part by NSF Grant No. PHY-0653656.


\begin{thebibliography}{99}

\bibitem{bib:DFS}
  M.~Dine, W.~Fischler and M.~Srednicki,
  Nucl.\ Phys.\  B {\bf 189}, 575 (1981).

\bibitem{bib:DR81}
  S.~Dimopoulos and S.~Raby,
  Nucl.\ Phys.\  B {\bf 192}, 353 (1981).

\bibitem{bib:DF}
  M.~Dine and W.~Fischler,
  Phys.\ Lett.\  B {\bf 110}, 227 (1982).

\bibitem{bib:NO}
  C.~R.~Nappi and B.~A.~Ovrut,
  Phys.\ Lett.\  B {\bf 113}, 175 (1982).

\bibitem{bib:A-GCW}
  L.~Alvarez-Gaume, M.~Claudson and M.~B.~Wise,
  Nucl.\ Phys.\  B {\bf 207}, 96 (1982).

\bibitem{bib:DR83}
  S.~Dimopoulos and S.~Raby,
  Nucl.\ Phys.\  B {\bf 219}, 479 (1983).

\bibitem{bib:DN}
  M.~Dine and A.~E.~Nelson,
  Phys.\ Rev.\  D {\bf 48}, 1277 (1993)
  [arXiv:hep-ph/9303230].

\bibitem{bib:DNS}
  M.~Dine, A.~E.~Nelson and Y.~Shirman,
  Phys.\ Rev.\  D {\bf 51}, 1362 (1995)
  [arXiv:hep-ph/9408384].

\bibitem{bib:DNNS}
  M.~Dine, A.~E.~Nelson, Y.~Nir and Y.~Shirman,
  Phys.\ Rev.\  D {\bf 53}, 2658 (1996)
  [arXiv:hep-ph/9507378].

\bibitem{bib:GR}
  G.~F.~Giudice and R.~Rattazzi,
  Phys.\ Rept.\  {\bf 322}, 419 (1999)
  [arXiv:hep-ph/9801271].

\bibitem{Izawa:1997gs}
  K.~I.~Izawa, Y.~Nomura, K.~Tobe and T.~Yanagida,
  Phys.\ Rev.\  D {\bf 56}, 2886 (1997)
  [arXiv:hep-ph/9705228].

\bibitem{Csaki:2006wi}
  C.~Csaki, Y.~Shirman and J.~Terning,
  JHEP {\bf 0705}, 099 (2007)
  [arXiv:hep-ph/0612241].

\bibitem{Dine:2007dz}
  M.~Dine and J.~D.~Mason,
  arXiv:0712.1355 [hep-ph].

\bibitem{Terning:1999at}
  J.~Terning and M.~A.~Luty,
  arXiv:hep-ph/9903393.

\bibitem{bib:CRS}
  A.~G.~Cohen, T.~S.~Roy and M.~Schmaltz,
  JHEP {\bf 0702}, 027 (2007)
  [arXiv:hep-ph/0612100].

\bibitem{bib:RS}
  T.~S.~Roy and M.~Schmaltz,
  Phys.\ Rev.\  D {\bf 77}, 095008 (2008)
  [arXiv:0708.3593 [hep-ph]].

\bibitem{bib:MSS}
  P.~Meade, N.~Seiberg and D.~Shih,
  arXiv:0801.3278 [hep-ph].

\bibitem{bib:CDFM}
  L.~M.~Carpenter, M.~Dine, G.~Festuccia and J.~D.~Mason,
  arXiv:0805.2944 [hep-ph].

\bibitem{Buican:2008ws}
  M.~Buican, P.~Meade, N.~Seiberg and D.~Shih,
  arXiv:0812.3668 [hep-ph].

\bibitem{bib:IS}
  K.~A.~Intriligator and M.~Sudano,
  arXiv:0807.3942 [hep-ph].

\bibitem{bib:Carpenter}
  L.~M.~Carpenter,
  arXiv:0809.0026 [hep-ph].

\bibitem{bib:SVW}
  N.~Seiberg, T.~Volansky and B.~Wecht,
  arXiv:0809.4437 [hep-ph].

\bibitem{bib:CFNV}
  C.~Csaki, A.~Falkowski, Y.~Nomura and T.~Volansky,
  arXiv:0809.4492 [hep-ph].

\bibitem{Komargodski:2008ax}
  Z.~Komargodski and N.~Seiberg,
  arXiv:0812.3900 [hep-ph].

\bibitem{Marques:2009yu}
  D.~Marques,
  arXiv:0901.1326 [hep-ph].

\bibitem{Martin:1996zb}
  S.~P.~Martin,
  Phys.\ Rev.\  D {\bf 55}, 3177 (1997)
  [arXiv:hep-ph/9608224].

\bibitem{Dimopoulos:1996gy}
  S.~Dimopoulos, G.~F.~Giudice and A.~Pomarol,
  Phys.\ Lett.\  B {\bf 389}, 37 (1996)
  [arXiv:hep-ph/9607225].

\bibitem{bib:Softsusy}
  B.~C.~Allanach,
  Comput.\ Phys.\ Commun.\  {\bf 143}, 305 (2002)
  [arXiv:hep-ph/0104145].

\bibitem{bib:PDG}
  C.~Amsler {\it et al.}  [Particle Data Group],
  Phys.\ Lett.\  B {\bf 667}, 1 (2008).

\bibitem{bib:BCG}
  H.~Baer, K.~m.~Cheung and J.~F.~Gunion,
  Phys.\ Rev.\  D {\bf 59}, 075002 (1999)
  [arXiv:hep-ph/9806361].

\bibitem{bib:CLNP}
  Z.~Chacko, M.~A.~Luty, A.~E.~Nelson and E.~Ponton,
  JHEP {\bf 0001}, 003 (2000)
  [arXiv:hep-ph/9911323].

\bibitem{bib:KKS}
  D.~E.~Kaplan, G.~D.~Kribs and M.~Schmaltz,
  Phys.\ Rev.\  D {\bf 62}, 035010 (2000)
  [arXiv:hep-ph/9911293].

\bibitem{bib:CKSS}
  H.~C.~Cheng, D.~E.~Kaplan, M.~Schmaltz and W.~Skiba,
  Phys.\ Lett.\  B {\bf 515}, 395 (2001)
  [arXiv:hep-ph/0106098].

\end{thebibliography}
 \end{document}